\documentclass[sigconf]{acmart}
\AtBeginDocument{%
  }

\copyrightyear{2026}
\acmYear{2026}
\acmConference{Poster at Graphics Interface 2026 (GI '26)}{June 9--12, 2026}{Waterloo, ON, Canada}

\begin{document}

\title{"A Glimpse, Not a Gaze": Using Generative AI to Balance Privacy and Awareness in Inter-generational Caregiving}

\author{Zixi Christina Li}
\orcid{0009-0006-4924-5780}
\affiliation{
\institution{University of Waterloo}
  \city{Waterloo}
  \state{Ontario}
  \country{Canada}}
\email{christina.li1@uwaterloo.ca}

\author{Keiko Katsuragawa}
\orcid{0000-0002-9642-9666}
\affiliation{
\institution{National Research Council Canada}
\institution{University of Waterloo}
  \city{Waterloo}
  \state{Ontario}
  \country{Canada}}
\email{keiko.katsuragawa@nrc-cnrc.gc.ca}

\author{James R. Wallace}
\orcid{0000-0002-5162-0256}
\affiliation{
\institution{University of Waterloo}
  \city{Waterloo}
  \state{Ontario}
  \country{Canada}}
\email{james.wallace@uwaterloo.ca}

\renewcommand{\shortauthors}{Trovato et al.}

\begin{abstract}
As older adults increasingly prefer to age in place, their adult children often assume the role of informal caregivers. This dynamic creates a distinct tension between the adult child's need for awareness and the older adult's fundamental right to privacy. Traditional monitoring technologies, such as raw video feeds, often compromise the older adult's autonomy. To address this challenge, this study explores the use of generative Artificial Intelligence (GenAI) to create abstract, privacy-preserving ``visual summaries'' of daily activities. We design a 10-day Experience Sampling Method (ESM) study with dyads consisting of older adults and their adult children. Through daily smartphone prompts, participants report their current context and evaluate pre-generated AI sketches, indicating their willingness to share or receive these images. Follow-up interviews will further investigate participants' boundary-setting behaviours. This research aims to quantify the privacy mismatch between generations and provide actionable design guidelines for applying visual abstraction in AI-mediated caregiving tools, ultimately supporting inter-generational connection while protecting user dignity.
\end{abstract}



\maketitle

\section{Introduction}
Older adults nowadays express a strong preference for aging in place, defined as the ability to remain living in one's own home and community safely, independently, and comfortably as one grows older \cite{wiles2012meaning}. While this approach significantly supports emotional well-being and autonomy, it often requires adult children to adopt the role of informal caregivers to assist from a distance \cite{homaeian2025}. To maintain awareness and peace of mind, many adult children turn to communication and monitoring technologies. However, traditional systems like raw video feeds or continuous sensor tracking often create a direct conflict between the adult child's need for awareness and the older adult's fundamental right to privacy \cite{li_privacy_2023}. 
To address this tension, this study explores the use of GenAI to transform daily activity data into abstract, privacy-preserving visual summaries. By evaluating how these AI-generated sketches are perceived, we aim to establish design guidelines that balance inter-generational connection while preserving the personal boundaries.

\section{Background}
As older adults increasingly choose to ``age-in-place'' \cite{krause2007longitudinal}, adult children frequently take on the complex role of informal caregivers~\cite{homaeian2025, lindley2008designing}. A central challenge in this dynamic is the inherent tension between an adult child's desire to monitor their parent for safety and the older adult's fundamental need for privacy \cite{li_privacy_2023}. Because traditional monitoring systems, such as raw video feeds, often compromise the autonomy and dignity of the older adult, there is a critical need for technologies that allow both parties to negotiate and establish firm boundaries around shared in-home activity data~\cite{homaeian2025, li_privacy_2023}.

To navigate the tension between privacy and awareness, researchers have explored abstracting personal data through Casual Information Visualization \cite{pousman_casual_2007}. GenAI has recently emerged as a powerful tool in this space, allowing systems to transform sensitive, raw data into engaging, non-photorealistic visual forms that foster reflection and meaning-making \cite{park_reimagining_2025, yun_exploring_2025}. By leveraging techniques like ``privacy avatars'' to mask literal data while still communicating user status \cite{swienty_investigating_2025}, GenAI visual summaries can successfully provide adult children with the ambient awareness they desire while applying necessary visual abstraction to protect the older adult's privacy~\cite{li_privacy_2023}.

\section{Research Questions}
\begin{itemize}
    \item RQ1: What specific aspects of everyday life do aging parents and adult children desire to see in a visual summary, and which latent details — those they might not explicitly ask for — emerge as most meaningful for connection?
    \item RQ2: What aspects do older adults and adult children consider inappropriate to include in AI-generated activity summaries shown to the other party?
\end{itemize}

\section{Methods}

The study recruits participant pairs, consisting of one older adult aged 65 or older who lives independently and one of their adult children aged 18 to 64. Participants must live in separate households and actively communicate with each other. All participants must be fluent in English, and own a smartphone with active internet access (data plan or consistent Wi-Fi) that they use daily. The research begins with a baseline survey where participants report their communication habits, occupation and daily activities, and they should establish a personalized visual avatar by selecting features such as facial appearance and hairstyle.


Based on the responses of this baseline survey from the participants, the researchers will use GenAI to create a simple, abstract sketch of their daily activities with their possible availability to connect with their children/parents. These images are pre-loaded into the Qualtrics library and categorized to match specific combinations of participant responses (See Figure 1).

\begin{figure}[h!]
\centering
\includegraphics[width=0.8\linewidth]{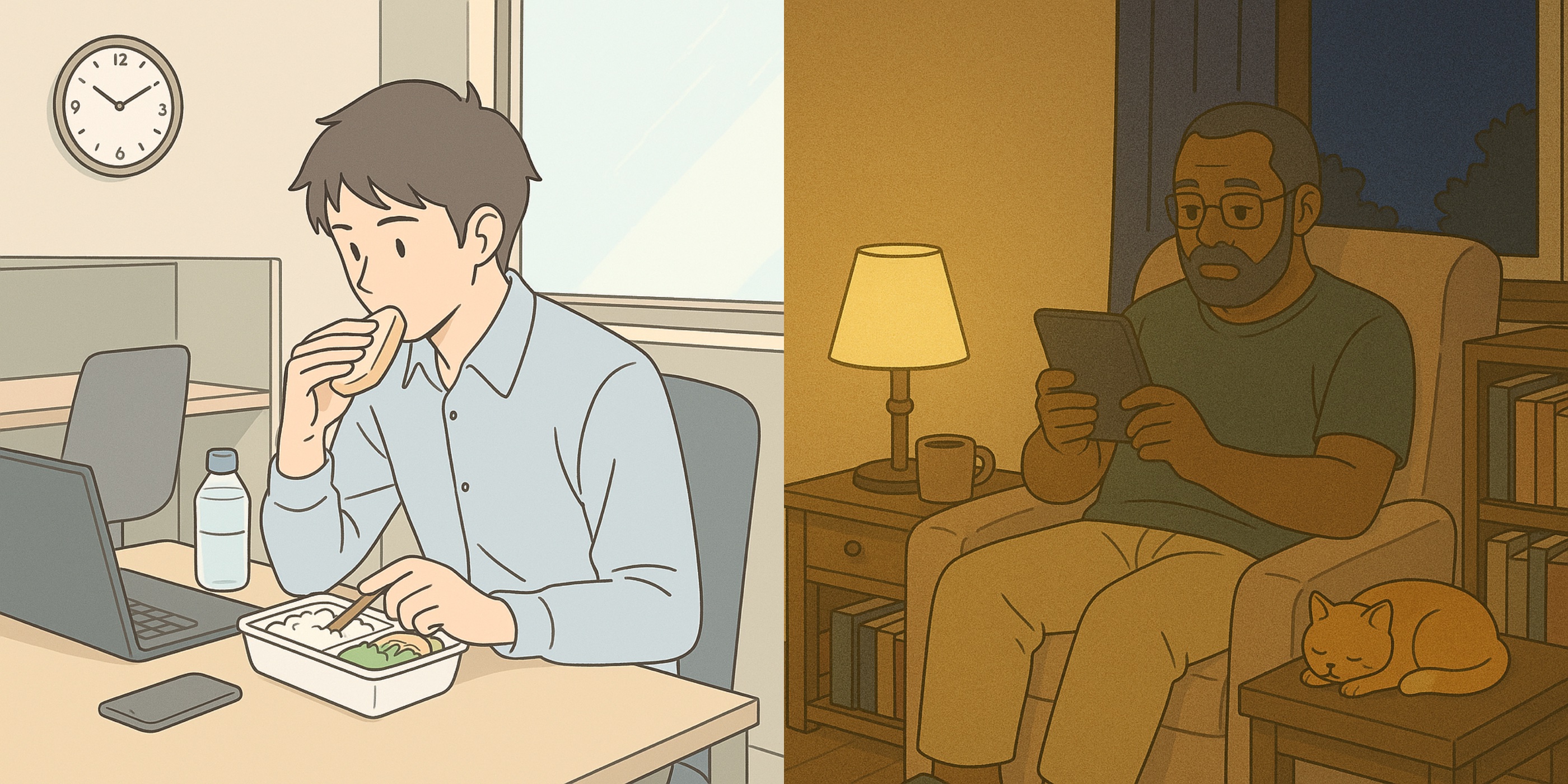}
\caption{Sample of AI generated sketches of participants}

\end{figure}

Next, participants enter a 10-day ESM phase. ESM is a validated approach that prompts participants in their natural environments, capturing \textit{in situ} experiences and significantly reducing the recall bias associated with traditional retrospective surveys \cite{gross2023experience}. During this time, they will receive three to four secure Qualtrics\cite{qualtrics2026} survey links via email on their preferred device each day. Through these survey links, participants can report their current activity, mood, and level of availability. Based on these inputs, the survey logic directs participants to a corresponding pre-generated, abstract sketch of their situation. Participants then evaluate whether they are willing to share this specific image with their family member, and whether they would want to receive a similar image from them. 

After the 14-day period ends, the researchers will schedule an interview with each participant individually. Prior to these sessions, the researchers will aggregate the images generated during the participant's ESM phase and prepare them for a "Card Sorting" activity. During the interview, participants will review these specific images and sort them into categories of appropriateness (e.g., "Happy to Share" vs. "Never Share"). The interview session will also involve a semi-structured discussion where participants explain the rationale behind their privacy boundaries, define their preferences for visual abstraction (e.g., blurring vs. masking), and discuss any behavioural changes they experienced, such as feeling more connected or worrying less due to seeing the visual summaries, allowing the research team to explore their privacy boundaries and design preferences.

\section{Expected Contributions}

This research will quantify and qualify the privacy and boundary concerns and expectations. By identifying these patterns, the study provides concrete insights into how family members navigate and set privacy boundaries. Ultimately, understanding these dynamics will inform the design of better inter-generational communication tools that satisfy the needs of both parties, directly supporting older adults who choose to age-in-place.

Additionally, the study will provide practical design guidelines for using visual abstraction to protect privacy. It will evaluate specific ways to hide sensitive information, such as blurring a background or completely removing certain objects from an image. By showing how to successfully manage what gets shared, these findings will help developers build AI caregiving tools that respect the older adults' and adult children's dignity and independence.

\bibliographystyle{ACM-Reference-Format}
\bibliography{sample-sigconf-authordraft}
\end{document}